\documentclass[a4paper,11pt]{article}
\pdfoutput=1 

\usepackage{lineno}
\usepackage{adjustbox}
\usepackage{subfigure}
\usepackage{multirow}
\usepackage{color}
\usepackage{url}
\usepackage{authblk}

\title{\boldmath CEPC Research Report: Higgs Physics Analysis}

\author[a]{Manqi Ruan}
\author[a]{Yaquan Fang}
\author[a]{Gang Li}
\author[a]{Dan Yu}
\affil{for the CEPC Physics-Detector Study Group}
\affil[a]{Institute of High Energy Physics, Chinese Academy of Science, Beijing, China}

\date{}
\begin{document}
\maketitle

\abstract{

In November 2018, the CEPC released the Conceptual Design Report (CDR) with the physics potential investigation and the analysis about different detector concepts.
Since then, intensive researches on the Higgs Physics at the CEPC have been progressed.
The run at top thresholds are also taken into consideration and the physics potential around the center-of-mass energy of 360GeV is studied.
The $ttH$ channel information boost the precision of Higgs width measurement by a factor of 2 with respect to 240 GeV Higgs Runs.
Different Higgs decay modes are analyzed in the CEPC Higgs factory at the center-of-mass energy of 240GeV, including Higgs to b/c/g and Higgs to $\tau\tau$, which improves the Higgs measurement.
Advanced analysis technologies are developed and applied in the analysis, such as jet reconstruction and lepton identification in jets.
New interpretations are under developing to improve the performance.
}

\newpage

The Circular Electron Positron Collider (CEPC) was proposed after the Higgs boson discovery at the Large Hadron Collider (LHC). 
Compared with the LHC, electron-positron colliders have significant advantages for the Higgs boson property measurements. 
The Higgs boson signal is largely free of QCD backgrounds and the signal to background ratio is significantly higher. 
Moreover, Higgs boson candidates can be identified through the recoil mass method without tagging the Higgs boson decay products, allowing for the measurements of the Higgs boson width and couplings in a model-independent way.
With about one million Higgs bosons produced at the CEPC, many of the major Higgs boson couplings can be measured with precisions about one order of magnitude better than those achievable at the High Luminosity-LHC.

The CEPC Conceptual Design Report (CDR) was released\cite{CEPCStudyGroup:2018ghi} in November 2018.
In the CEPC-CDR, different detector concepts have been proposed.
These concepts can fulfill the physics requirements, confirmed by the dedicated simulation and R\&D program.
The joint theory, detector design, and simulation studies have given a clear demonstration of the CEPC physics capabilities: the CEPC has an immense potential towards the precision Higgs and EW measurements, and can provide complementary information in flavor physics.

More Higgs studies have been progressed after the CEPC-CDR delivery, including run at top thresholds, differential measurements, new analysis technologies, and new interpretations.

\section{Run at top thresholds}

The $ttH$ channel information boost the precision of Higgs width measurement by a factor of 2 with respect to 240 GeV Higgs Runs\cite{Kaili:yangzhou}. 
The Table \ref{precisionTab} shows the combination of CEPC Higgs $\sigma \times Br$ and coupling measurement precision. 

\begin{table}[htbp]
\centering
\caption{\label{precisionTab}CEPC Higgs measurement precision} 
\smallskip
\begin{tabular}{ccccc}
\hline
& 240GeV, 5.6$ab^{-1}$&  \multicolumn{2}{c}{ 360GeV, 2$ab^{-1}$}\\
\hline
& ZH & ZH & $\nu\nu$H \\
any & 0.50\% & 1\% & - \\
$H\to bb$ & 0.27\% & 0.63\% & 0.76\%  \\
$H\to cc$ & 3.3\% & 6.2\% & 11\%  \\
$H\to gg$ & 1.3\% & 2.4\% & 3.2\%  \\
$H\to WW$ & 1.0\% & 2.0\% & 3.1\%  \\
$H\to ZZ$ & 7.9\% & 14\% & 15\%  \\
$H\to \tau\tau$ & 0.8\% & 1.5\% & 3\%  \\
$H\to \gamma\gamma$ & 5.7\% & 8\% & 11\%  \\
$H\to \mu\mu$ & 12\% & 29\% & 40\%  \\
$Br_{upper}$($H\to inv.$) & 0.2\% & - & -  \\
$\sigma(ZH)\times BR(H\to Z\gamma)$ & 16\% & 25\% & - \\
width & 2.9\% &  \multicolumn{2}{c}{ 1.4\%}   \\

\hline
\end{tabular}
\end{table}

The couplings of Higgs to top quark are the key driver of the hierarchy problem.
The CEPC sensitivity to Higgs and top Effective Field Theory (EFT) is analyzed using the minimal Higgs Top anomalous coupling EFT set\cite{topcp}.
The $tt$ threshold scan in CEPC can measure the top mass at a precision 1 order of magnitude better than hadron colliders, estimated by constructing 1-dimension likelihood combining the statistical power of all scan points\cite{topmass}.
Besides, it provides the simultaneous measurement of top width, top Yukawa coupling and the $\alpha_{s}$.
A few proposals of luminosities and energies are also tested.

\section{The di-Higgs related physics}

\begin{figure}[htbp]
\centering
\includegraphics[width=.55 \textwidth,clip]{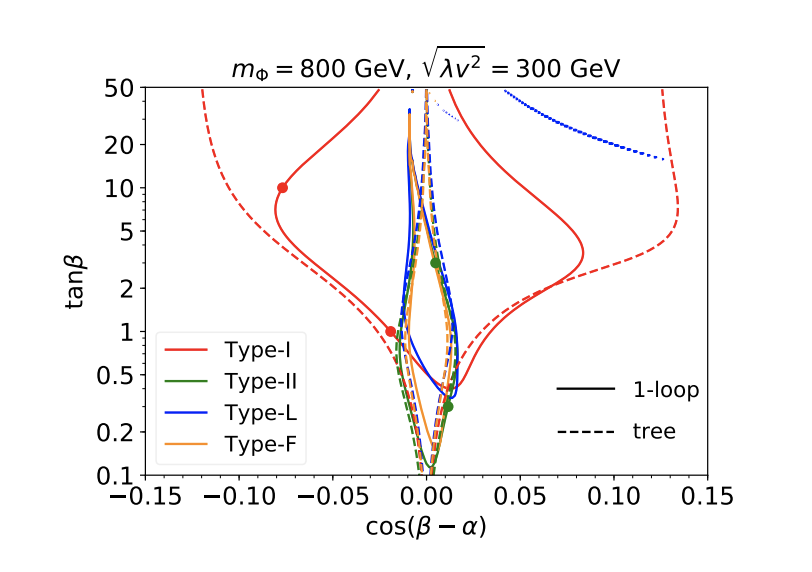} 
\label{cont_2HDM}
\caption{5$\sigma$ discovery regions outside the contour lines in the $\cos(\beta -\alpha)$-$\tan \beta$ plane for CEPC at tree level (dashed) and one-loop (solid). 
Red, green, blue and orange colors indicate the results of Type-I, II, L and F, respectively. 
We choose $m_{H}=m_{A}=m_{H^{\pm}}=800GeV$ and $\sqrt{\lambda \nu^{2}}=300GeV$ for one-loop curves. 
}
\end{figure}

The studying of different two Higgs doublet model (2HDM) illustrates that the future precision measurements of the Standard Model (SM) observables at the proposed Z-factories and Higgs factories may have impacts on new physics beyond the Standard Model.
The achievable sensitivity of the CEPC to probe the virtual effects of the 2HDM, including Type-II\cite{Chen:2018shg}, Type-I\cite{Chen:2019pkq}, Type-L (lepton specific) and Type-F (flipped Yukawa couplings)\cite{Han:2020lta}, has been examined.
The regions outside the contour lines in Figure \ref{cont_2HDM} are accessible with 5$\sigma$ sensitivity. 
The discovery potential is explored based on the hypothetical deviations in the precision data for the 2HDMs up to one-loop level.
A large part of the parameter space of the other types of 2HDMs can be distinguished from the benchmark points of the target model. 
The impacts of loop corrections are found to be significant in certain parameter regions.
 
\section{The measurement of the Higgs boson decaying to fermions}
With the combination of detector hardwares and multivariable analysis algorithms, the measurement efficiency of charm quark and bottom quark jet is improved.
Therefore, more effective evidence of the coupling between Higgs boson and the first two generations of quarks can be found, and the coupling strength between Higgs and the third generation heavy quarks can be measured more accurately.
The measurement of branching fraction of $H\to bb/cc/gg$ are studied in $\mu\mu H$ and $eeH$ process\cite{Bai:2019qwd}, .
The statistical uncertainty of the signal cross section is estimated to be about 1\% in the $H\to bb$ final state, and approximately 5\%-10\% in the $H\to cc/gg$ final states. 
The systematic uncertainties on the branching fraction measurements are also studied, which are around 0.6\%, 6\% and 8\% for $bb, cc$ and $gg$ final states respectively.
The high precision of this measurement benefits from the distinct signature of events with the Higgs boson and clean background in electron-positron collider, as well as the model independent analysis.
This study demonstrates the potential of precise measurement of the hadronic final states of the Higgs boson decay at the CEPC, and will provide key information to understand the Yukawa couplings between the Higgs boson and quarks, which are predicted to be the origin of quarks' masses in the standard model.
\begin{figure}[htbp]
\centering
\includegraphics[width=.9 \textwidth,clip]{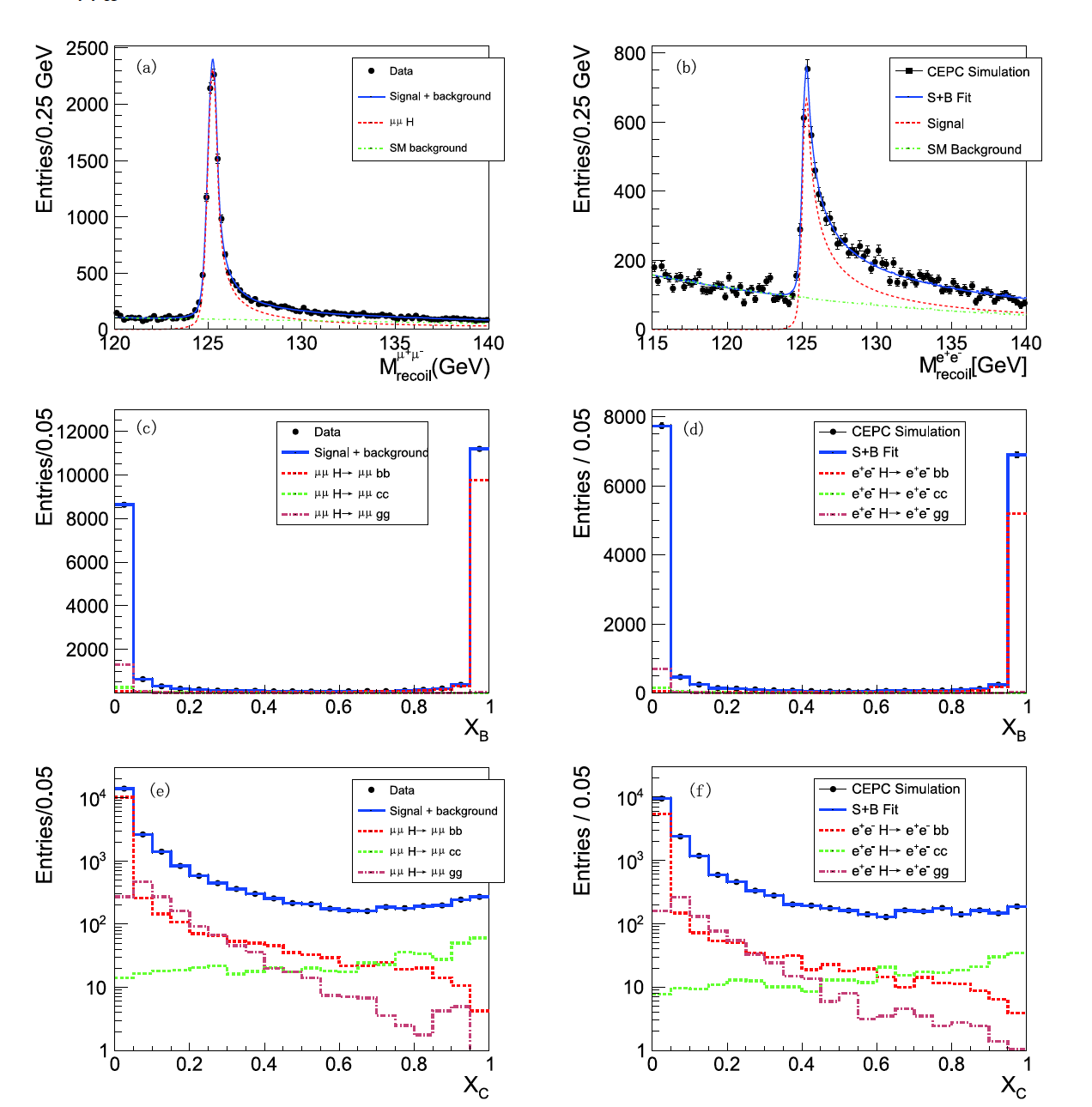} 
\label{Higgsbcg}
\caption{3D-fit result projected on recoil mass distribution in $\mu\mu H$ (a) channel and $eeH$ channel (b), on B-likeness distribution in $\mu\mu H$ (c) channel and $eeH$ channel (d), and on C-likeness distribution in $\mu\mu H$ (e) channel and $eeH$ channel (f)}
\end{figure}

\section{Precise Higgs measurement in CEPC}
By combining the analysis of different channels, the advantage of CEPC in Higgs property measurement accuracy can be fully exploited, and the physics potential can be studied quantitatively. 
The key physics measurement is improved with innovative reconstruction methods and improved signal-background discrimination.
The Higgs invisible decay and $\tau$ decay, are studied and corresponding articles have been published.
Moreover, important studies relevant to the Higgs are processed, including jet lepton identification, jet reconstruction, and combined analysis technology.

\subsection{$H\to invisible$ channel}
\begin{figure}[htbp]
\centering
\includegraphics[width=.55 \textwidth,clip]{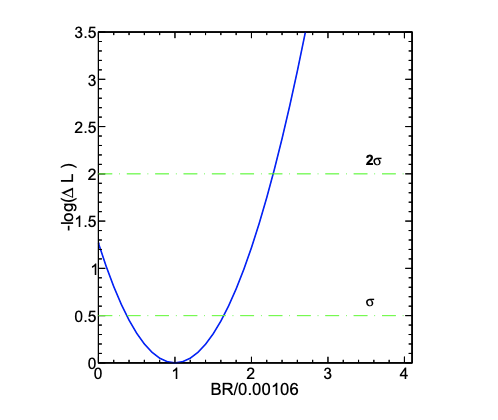} 
\label{Higgsinv}
\caption{The likelihood profile of combination, where the green projective line label out the location of 68\%, 95\% confidence level, which corresponds to $-\Delta \log(L)=0.5,2 on the y-axis$.}
\end{figure}

The measurement potential of Higgs decay to an invisible channel on CEPC is investigated\cite{Tan:2020ufz}.
In SM, the Higgs boson can only decay invisibly via $H\to ZZ^{*} \to \nu\nu\nu\nu$ or dark matter, so any evidence of invisible Higgs decay that exceeds BR ($H\to inv$.) will immediately point to a phenomenon that is beyond the standard model (BSM). 
In this analysis, the upper limit of BR ($H\to inv$) is estimated for three channels, including two leptonic channels and one hadronic channel, under the assumption predicted by SM. 
With the SM ZH production rate, the upper limit of BR ($H\to inv$) could reach 0.24\% at the 95\% confidence level, as shown in Figure\ref{Higgsinv}.
The impact of the BMR on the ZH ($Z\to qq, H\to inv$.) signal strength accuracy has been studied as well, and the obtained curve shows that the analysis result would be one of the indicators for the CEPC detector optimization study.

\subsection{$H\to \tau\tau$ channel}
\begin{figure}[htbp]
\centering
\includegraphics[width=.5 \textwidth,clip]{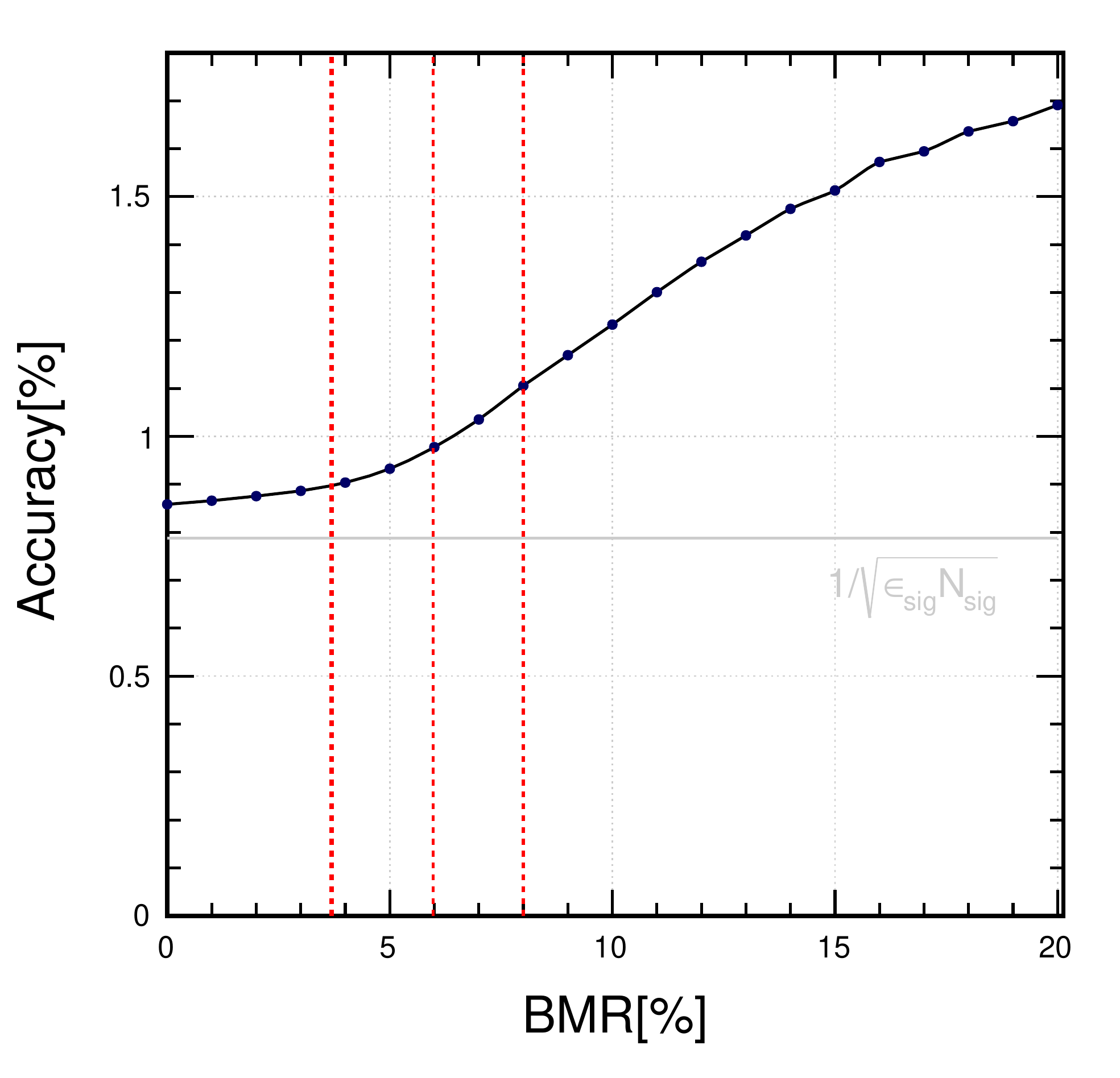} 
\label{Higgstau}
\caption{The dependence of the accuracy of the qqH channel on the boson mass resolution. The thicker red line indicates that the CEPC baseline BMR is 3.8\%, the other two red lines show the accuracy when the BMR drops to 6\% and 8\%. (Signal: $ZH \to qq\tau\tau$, Background: $ZZ \to qq\tau\tau$,)}
\end{figure}

In the $H \to\tau\tau$ analysis, the signal strength measurement is studied\cite{Yu:2020bxh}.
Using the full simulation analysis, the CEPC is expected to measure the signal strength to a relative accuracy of 0.8\%. 
Multiple analysis technologies and dedicated $\tau$ finding algorithms are developed.
A precise reconstruction of the impact parameter is essential for the $\tau$ events identification, and the PFA oriented detector design and reconstruction are critical for this analysis.
The physics requirement on the mass resolution of the Higgs boson with hadronic decay final states is also discussed, as shown in Figure \ref{Higgstau}.
It shows that the CEPC baseline design and reconstruction fulfill the accuracy requirement of the $H\to\tau\tau$ signal strength. 

\subsection{Jet reconstruction}
\begin{figure}[htbp]
\centering
\subfigure{
\includegraphics[width=.3 \textwidth,clip]{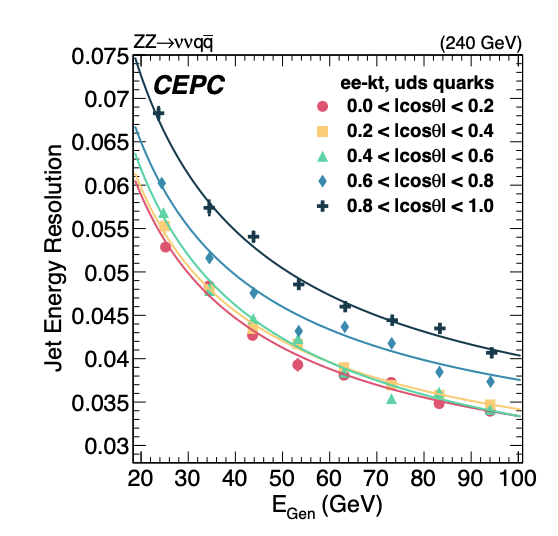} 
\label{jetreco_jer}
}
\subfigure{
\includegraphics[width=.3 \textwidth,clip]{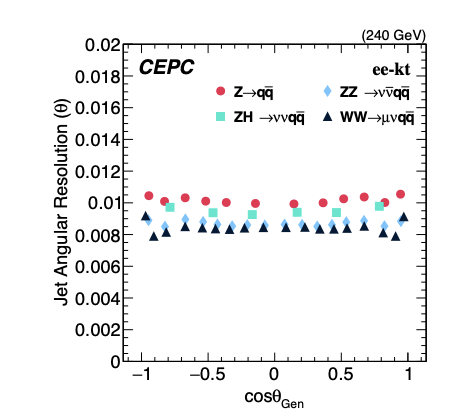} 
\label{jetreco_jar1}
}
\subfigure{
\includegraphics[width=.3 \textwidth,clip]{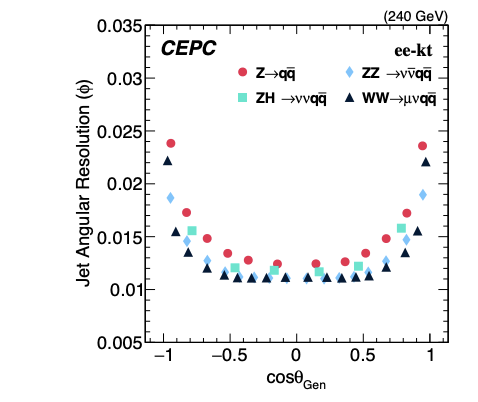} 
\label{jetreco_jar2}
}
\label{jetreco}
\caption{(a) is the JER as a function of the GenJet energy of light-flavored (uds) jets from semi-leptonic decay are divided into five fiducial regions. (b-c) is the polar and azimuth angular resolutions as a function of the GenJet $\cos\theta$ in several benchmark 2-jet events.}
\end{figure}

Jet reconstruction is critical for the precision measurement of Higgs boson properties and the electroweak observables at the CEPC\cite{Lai:2021rko}.
The jet energy and angular responses of benchmark 2- and 4-jet processes are analyzed with fully simulated samples with the CEPC baseline detector geometry. 
As shown in Figure \ref{jetreco}, the relative resolution of 3.5\% and 1\% are observed on the jet energy and angular measurement for jets in the detector barrel region ($\left| \cos \theta \right| < 0.6$) with energy greater than 60 GeV. 
Meanwhile, the jet energy/angular scale can be controlled within 0.5/0.01\%. 
The differential dependences of the jet response on the jet direction and energy are extracted. 
The impact on the jet responses induced by different jet clustering algorithms and matching criteria are also analyzed, which yields a relative difference of up to 8\%.

\subsection{Jet lepton identification}
\begin{figure}[htbp]
\centering
\subfigure{
\includegraphics[width=.4 \textwidth,clip]{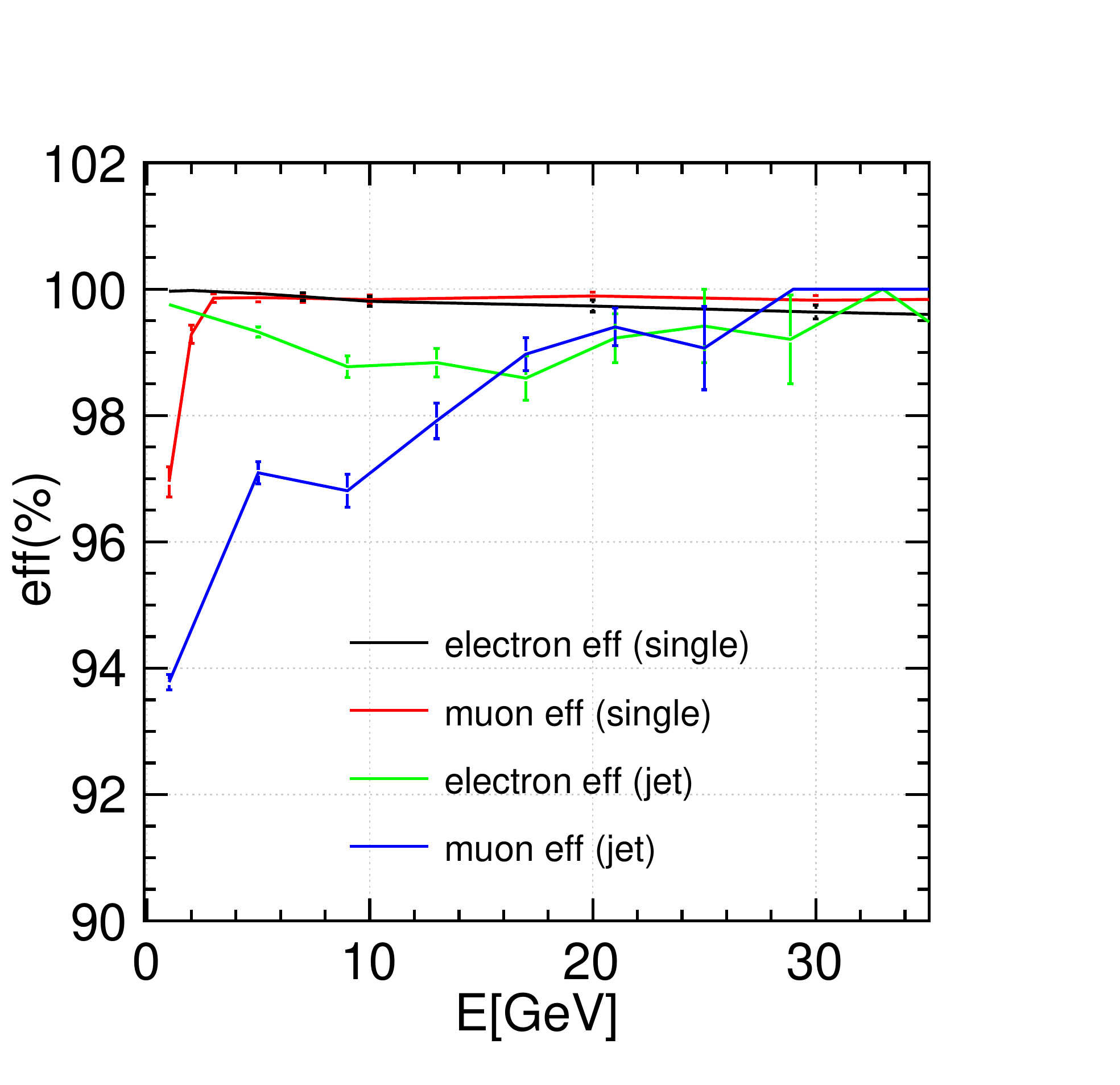} 
\label{Compare_all_eff}
}
\subfigure{
\includegraphics[width=.4 \textwidth,clip]{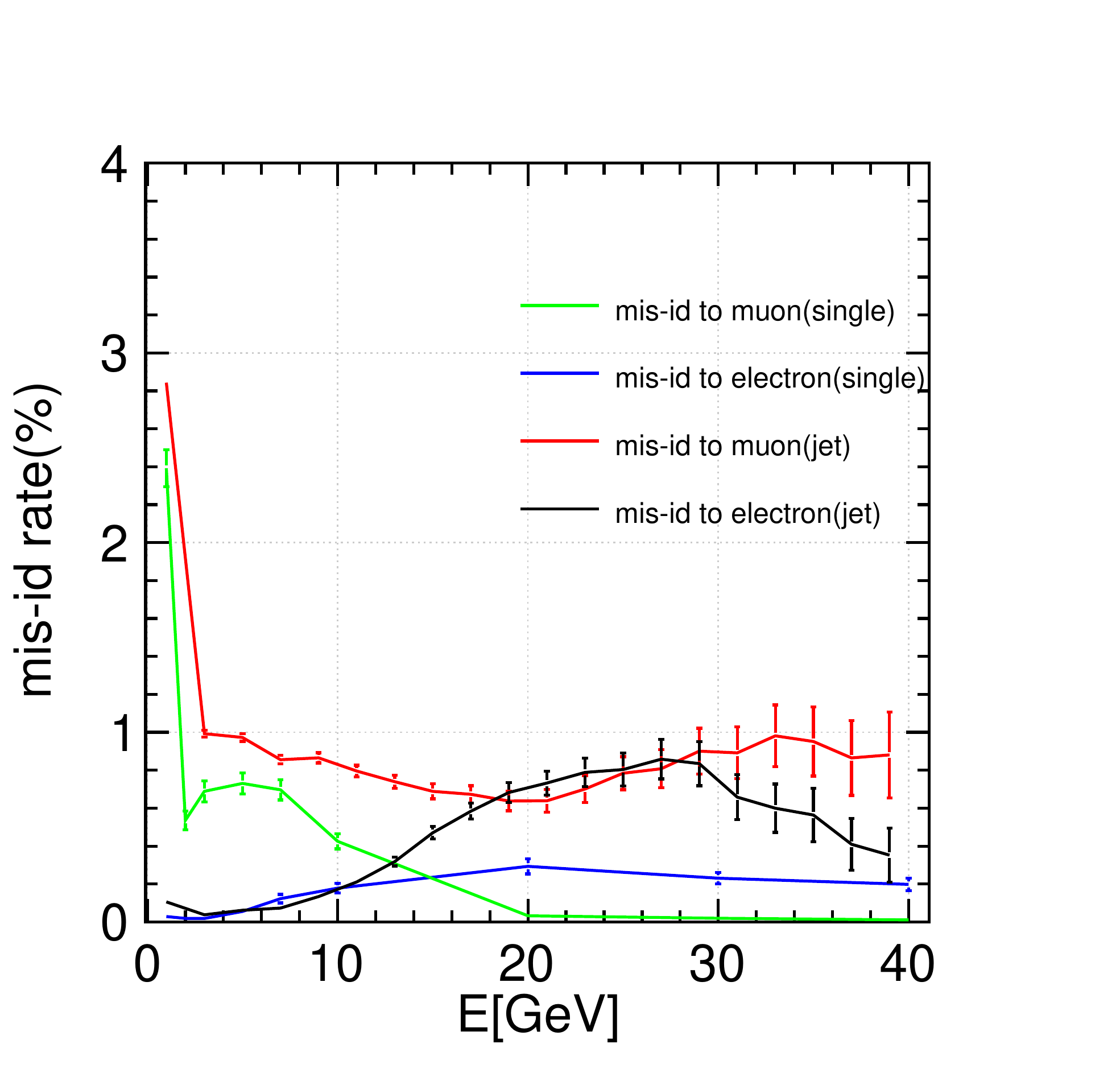} 
\label{Compare_all_mis}
}
\label{Compare_all}
\caption{The efficiency and purity of electrons and muons in jets, comparing with single particle case.}
\end{figure}

Identifying the leptons inside jets is critical for the measurements of Higgs boson and the flavor physics program at the CEPC\cite{jetlepton}.
Using the CEPC baseline software, the identification performance of leptons generated inside a jet is analyzed. 
The jet leptons are identified with typical efficiency and mis-identification rate of 98\% and 1\% for energy higher than 2 GeV, with the $Z\to bb$ process at 91.2 GeV center-of-mass energy. 
At the benchmark channel of the CEPC flavor program of $B_{c}\to \tau\nu$ with $\tau\to e\nu\nu$, the electrons are identified with inclusive efficiency times purity of 97\%, providing sufficient signal selection for the physics potential study. 
Compared to the single leptons, these jet leptons identification efficiency degrades about 1-3\%, and the mis-identification rate increases by less than 1\% at the same working point, as shown in Figure \ref{Compare_all}. 
The dependence between the performance of the shower reconstruction and the lepton identification is quantified.

\subsection{Global analysis}

A global analysis is proposed to improve the precision of a set of branching fractions of same resonance significantly, by taking all the cross talks among all processes into account and benefiting from statistical character of the multinomial distribution or binnomial distribution in the simplest case. It was first used in $\tau$ decay branching fractions by the ALEPH collaboration~\cite{Schael:2005am} and a good demonstration can also be found in~\cite{Ablikim:2014mww} by the BESIII collaboration.  Usually only one signal and its background processes are studied in a branching fraction measurement,  which is a  typical classification problem of signal and background processes.  If more than one decay modes from same resonance are measured simultaneously, taking the Higgs decay as an example, therefore two types of "backgrounds" must be studied, one is the backgrounds from other Standard Model (SM) processes, the other is the contamination among all the decay modes of the Higgs (cross talks). 

The main ingredients of the global analysis are determination of the efficiency matrices, matrix manipulation, and minimization of $\chi^2$.  The first one could be accomplished with ML technique elegantly~\cite{EFN,deepsets}. The last two are easily to be realized with some mature and widely used software tools. 

The improvement of the precision of the Higgs decay branching fractions brought by the global analysis approach could be explained with a few points.
First, the global analysis makes use of the full covariance of all counts, which means more information is used to determine the fitting parameters. Second, it deploys the law of multinomial distribution in the analysis. Therefore, the statistical uncertainties are significantly smaller than those of individual analysis, because they are calculated with produced numbers of signals according to the multinomial distribution and then are transformed into the ones of observed signals. Third, the global analysis introduces an unitary constraint of the total branching fractions. In conclusion, the global analysis approach adapts more information from data, statistical principle, and prior of physics. 

The preliminary results based on fast simulation  show that the precision could be improved by a factor of 2 in average~\cite{GLAN}.

\section{Conclusion}

The Higgs physics analysis at CEPC are progressing intensively. 
Different reconstruction methods of physics objects are developed, and different combinations of kinematic variables are applied in the analysis. 
More international collaborations are involved in the CEPC R\&D, and fruitful achievements are expected.

\end{document}